\documentclass[a4paper,11pt]{article}
\usepackage{pos}
\usepackage{tikz}
\usepackage{tikz-feynman}
\usepackage{braket}
\title{$V_{cs}$ determination from $D \to{}K \ell \nu$}

\author*[a,1]{W. G. Parrott}
\author[b,1]{Bipasha Chakraborty}
\author[a,1]{C. Bouchard}
\author[a,1]{C. T. H. Davies}
\author[c,1]{J. Koponen}
\author[d,1]{G. P. Lepage}

\affiliation[a]{SUPA, School of Physics and Astronomy, University of Glasgow, Glasgow, G12 8QQ, UK}
\affiliation[b]{DAMTP, Centre for Mathematical Sciences, University of Cambridge, Wilberforce Road, Cambridge, CB3 0WA}
\affiliation[c]{PRISMA+ Cluster of Excellence and Institut f\"{u}r Kernphysik, Johannes Gutenberg-Universit\"{a}t Mainz, D-55128 Mainz, Germany}
\affiliation[d]{Laboratory for Elementary-Particle Physics, Cornell University, Ithaca, NY 14853, USA}
\note{For the HPQCD collaboration}

\emailAdd{w.parrott.1@research.gla.ac.uk}
\emailAdd{bc335@cam.ac.uk}
\emailAdd{chris.bouchard@glasgow.ac.uk}
\emailAdd{christine.davies@glasgow.ac.uk}
\emailAdd{jonna.koponen@glasgow.ac.uk }
\emailAdd{gpl3@cornell.edu}

\abstract{Semileptonic $D \to{}K \ell \nu$ decays provide one angle of attack to get at the CKM matrix element $V_{cs}$, complementary to the study of leptonic $D_s$ decays. Here, HPQCD present the results of a recently published, improved determination of $V_{cs}$. 

We discuss a new, precise determination of $D\to K$ scalar and vector form factors from a lattice calculation on eight different $N_f=2+1+1$ MILC gluon field ensembles using the HISQ action, including three with physical light quark masses. When combined with experimental results, we are able to extract $|V_{cs}|=0.9663(80)$ to a sub percent level of precision for the first time. This is achieved using three different methods, which each combine our form factors with different sets of experimental results in different ways, with the results in very good agreement. Our primary method is to use $q^2$-binned data for the differential decay rate, but we also calculate $V_{cs}$ from the total branching fraction and from the value $|V_{cs}|f_+(0)$, which is also quoted by some experiments. }

\FullConference{%
 The 38th International Symposium on Lattice Field Theory, LATTICE2021
  26th-30th July, 2021
  Zoom/Gather@Massachusetts Institute of Technology
}

\begin{document}
\maketitle

\section{Introduction}
\begin{figure}
  \begin{center}
    \begin{tikzpicture}
      \begin{feynman}
        \vertex (a1) {\(\overline c\)};
        \vertex[right=2cm of a1] (a2);
        \vertex[right=0.5cm of a2] (a3);
        \vertex[above=0.2cm of a3] (a6){\(V_{cs}\)};
        \vertex[right=0.5cm of a3] (a4);
        \vertex[right=2cm of a4] (a5) {\(\overline s\)};

        \vertex[above=2cm of a1] (b1) {\(d\)};
        \vertex[above=2cm of a5] (b2) {\(d\)};
        \vertex[below=1.5em of a5] (c1) {\(\ell^-\)};
        \vertex[below=2.5em of a5] (c3) {\(\overline\nu\)};
        \vertex at ($(c1)!0.5!(c3) - (1cm, 0)$) (c2);
        \diagram* {
        {[edges=fermion]
        (a5) -- (a2) -- (a1),
        },
        (b1) -- [fermion] (b2),
        (c3) -- [fermion, out=180, in=-60] (c2) -- [fermion, out=60, in=180] (c1),
        (a3) -- [photon, bend right] (c2),
        };
      \end{feynman}
    \end{tikzpicture}
  \end{center}
  \caption{Feynman diagram for a $D^-\to K^0\ell^-\bar{\nu}$ decay.}
  \label{fig:DKfeyn}
\end{figure}
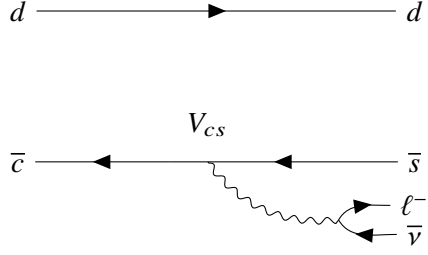

Flavour changing weak decays such as $D\to K\ell\nu$ can be used to test the Standard Model (SM). As depicted in Figure~\ref{fig:DKfeyn}, such decays involve Cabbibo-Kobayashi-Maskawa~\cite{Cabibbo:1963yz,Kobayashi:1973fv} (CKM) matrix elements, in this case $V_{cs}$. In the SM, the CKM matrix is unitary, and we can test this using independent determinations of the matrix elements. For the $D\to K$ semileptonic decay which we discuss here, we use lattice QCD to compute the hadronic form factors. These can then be combined with experimental data (details below) to determine $V_{cs}$. Because $V_{cs}\approx 1$, and other elements in the same row and column of the matrix are relatively small, it's important to know $V_{cs}$ very precisely in order to be able to carry out any meaningful unitarity test. Here, we summarise the findings in~\cite{Chakraborty:2021qav}, in which we make a significant improvement on the uncertainty in $V_{cs}$. For more detail, see~\cite{Chakraborty:2021qav}.
\section{Lattice calculation details}
\begin{figure}
  \begin{center}
    \hspace{-30pt}
    \includegraphics[width=0.5\textwidth]{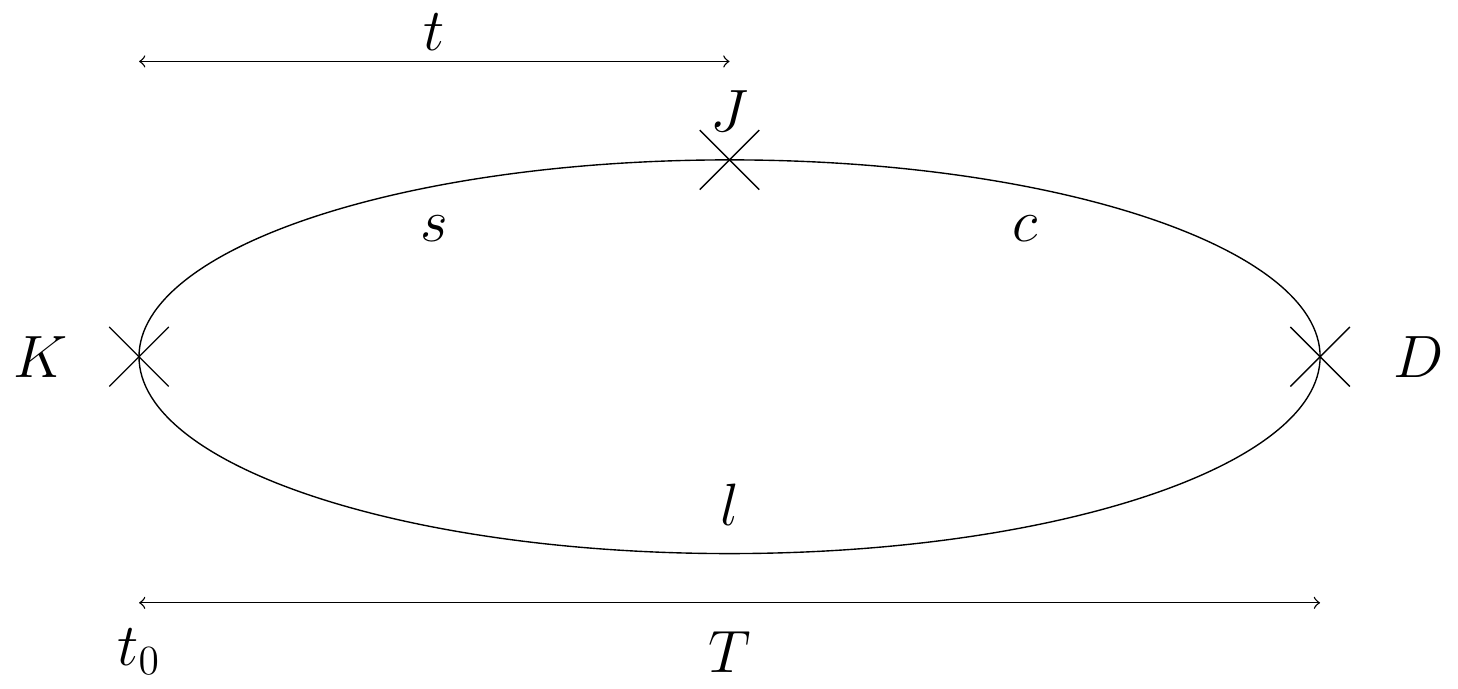}
    \caption{Schematic of a three-point correlation function for current insertion $J$. Note that we run the decay in reverse for computational convenience which does not affect the result.}
    \label{fig:decay}
  \end{center}
\end{figure}

We wish to calculate the scalar and vector ($f_0(q^2)$ and $f_+(q^2)$) form factors, as a function of the 4-momentum transfer squared $q^2=(p_D-p_K)^2$, over the full physical range, from zero recoil ($q^2_{\mathrm{max}}=(M_D-M_K)^2$) to maximum recoil ($q^2=0$). In order to calculate the form factors, we need matrix elements from three-point correlation functions, such as the one depicted schematically in Figure~\ref{fig:decay}. Setting up the calculation backwards for convenience, we start on time slice $t_0$, insert a current $J=S(V)$ for the scalar (vector) form factor at $t$ and finish on time slice $T$. We use $n_{\mathrm{src}}$ different $t_0$ values on each ensemble (see Table~\ref{tab:ensembles}), as well as 3-4 $T$ values. On each ensemble, we fit the correlation functions using a multi-exponential fit (see~\cite{Chakraborty:2021qav} for details) to extract the ground state amplitudes $J_{00}$.
Once we have $J_{00}$ from the three-point functions, we can convert them to matrix elements,
\begin{equation}\label{Eq:fitnormalisation}
  \bra{K}J\ket{D}=2\sqrt{M_{D}E_{K}}J_{00},
\end{equation}
and then combine them with masses from two-point functions to give the form factors,
\begin{equation}
  \begin{split} 
    Z_{V,t} \langle K|V^\mu|D\rangle =& f_+(q^2) [p_D^\mu + p_K^\mu - \frac{M^2_D - M^2_K}{q^2}q^\mu] + f_0(q^2) \frac{M^2_D - M^2_K}{q^2}q^\mu, \\
    \langle K|S|D\rangle =& \frac{M^2_D - M^2_K}{m_{c} - m_{s}}f_0^{D \rightarrow K}(q^2),
  \end{split}
\end{equation}
where meson masses are denoted with $M$s, quark masses with $m$s and the vector current is normalised non-perturbatively using the Partially Conserved Vector Current (PCVC) relation,
\begin{equation}
  Z_{V,t}\langle K|V^0|D\rangle|_{q^2_{\mathrm{max}}} = (M_D+M_K)f_0(q^2_{\mathrm{max}}).
\end{equation}
\begin{table*}
  \caption{Gluon ensembles used in this work, generated by the MILC collaboration~\cite{Bazavov:2012xda}. The Wilson flow parameter, $w_0=0.1715(9)\text{fm}$, is determined in~\cite{Dowdall:2013rya}, and is used to calculate the lattice spacing $a$ via values for $w_0/a$,~\cite{Borsanyi:2012zs} in column 3, which are from~\cite{McLean:2019qcx}. Column 4 gives the spatial ($N_x$) and temporal ($N_t$) dimensions of each lattice in lattice units and column 5 the number of configurations and time sources used in each case, whilst columns 6-10 give the masses of the valence and sea quarks, noting that these are the same in the case of the light quark.}
  \begin{center} 
    \begin{tabular}{c c c c c c c c c c}
      \hline
      Set & $\beta$ & $w_0/a$ & $N_x^3\times N_t$  &$n_{\mathrm{cfg}}\times n_{\mathrm{src}}$ &    $am_{l}^{\mathrm{sea/val}}$ & $am_{s}^{\mathrm{sea}}$ & $am_c^{\mathrm{sea}}$& $am_{s}^{\mathrm{val}}$ & $am_c^{\mathrm{val}}$\\
  \hline
  \hline
    1   & 5.8    & 1.1367(5)  &  $32^3\times 48$    & $998\times 16$ &     0.00235        &  0.0647        &  0.831&  0.0678 &  0.8605\\
    \hline
    2    & 6.0    & 1.4149(6)   &  $48^3\times 64$    & $985\times 16$&   0.00184       &   0.0507        &  0.628  &0.0527  & 0.643\\
    \hline
    3   &   6.3  & 1.9518(7)   & $64^3\times 96$    & $620\times 8$&  0.00120       &  0.0363         &  0.432&   0.036  &  0.433\\
    \hline
    4    & 5.8    & 1.1119(10) &  $16^3\times 48$    & $1020\times 16$&     0.013        &  0.065        &  0.838 &  0.0705 &  0.888 \\
    \hline
    5    & 6.0    &  1.3826(11) &  $24^3\times 64$   & $1053\times 16$ &     0.0102        &  0.0509        &  0.635  &  0.0545  & 0.664 \\
    \hline
    6    & 6.3    &  1.9006(20)  &  $32^3\times 96$  & $499\times 16$  &   0.0074       &   0.037        &  0.440&  0.0376  &  0.449\\
    \hline
    7   & 6.72    &  2.896(6)  & $48^3\times 144$    & $415\times 8$&  0.0048       &  0.024         &  0.286& 0.0234 &  0.274 \\
    \hline
    8   & 7.0    &   3.892(12) & $64^3\times 192$    & $375\times 4$&  0.00316       &  0.0158         &  0.188& 0.0165 &  0.194 \\
    \hline
    \hline
    \end{tabular}
  \end{center}
  \label{tab:ensembles}
\end{table*}
We calculate the necessary two-point and three-point correlation functions on 8 different  $N_f=2+1+1$ gluon ensembles from the MILC collaboration~\cite{Bazavov:2012xda}, detailed in Table~\ref{tab:ensembles}. Note that in our calculation light quarks are degenerate ($m_l=m_u=m_d$). Each ensembles has HISQ~\cite{Follana:2006rc} valence and sea quarks, and for three of them (sets 1, 2 and 3) the light quarks have physical masses. To span the $q^2$ range, several different momenta are imparted to the daughter quark on each ensemble using twisted boundary conditions.\\

In order to extrapolate to the continuum, we fit to a modified $z$ expansion, with $z=(\sqrt{t_+-q^2}-\sqrt{t_+})/(\sqrt{t_+-q^2}+\sqrt{t_+})$ and $t_+=(M_D+M_K)^2$. 
\begin{equation}\label{Eq:zexpansion}
  \begin{split}
    f_0(q^2)&=\frac{1+L}{1-\frac{q^2}{M^2_{D_{s}^{0}}}}\sum_{n=0}^{N-1}a_n^0z^n,\\
    f_+(q^2)&=\frac{1+L}{1-\frac{q^2}{M^2_{D_{s}^{*}}}}\sum_{n=0}^{N-1}a_n^+\Big(z^n-\frac{n}{N}(-1)^{n-N}z^N\Big),
  \end{split}
\end{equation}
where $L$ is a chiral logarithm term, to account for the light quark mass extrapolation and $M_{D_s^0}$ and $M_{D_s^*}$ are the masses of the scalar and vector $D_s$ respectively. $a_n^{0,+}$ is further broken down
\begin{equation}
  \begin{split}
    a_n^{0,+}&=\sum^{2}_{j=0}d_{jn}^{0,+}\Big(\frac{am_c^{\text{val}}}{\pi}\Big)^{2j}\times(1+\mathcal{N}^{0,+}_n),\\
     \mathcal{N}_n^{0,+}&=\frac{c_{s,n}^{\text{val},0,+}\delta_s^{\text{val}}+c_{l,n}^{\text{val},0,+}\delta_l^{\text{val}}+c_{s,n}^{0,+}\delta_s+2c_{l,n}^{0,+}\delta_l}{10m_s^{\text{tuned}}}+c_{c,n}^{0,+}\Big(\frac{M_{\eta_c}-M_{\eta_c}^{\text{phys}}}{M_{\eta_c}^{\text{phys}}}\Big),
  \end{split}
\end{equation}
to allow for discretisation effects and quark mass mistunings, with $d^{0,+}_{jn}$ and the $c$s tunable fit parameters. We obtain the continuum, physical point limit by setting $a=0$ and $\mathcal{N}_n^{0,+}=0$.
\section{Results}
\subsection{Form factors}
\begin{figure}
  \begin{center}
    \hspace{-30pt}
    \includegraphics[width=0.5\textwidth]{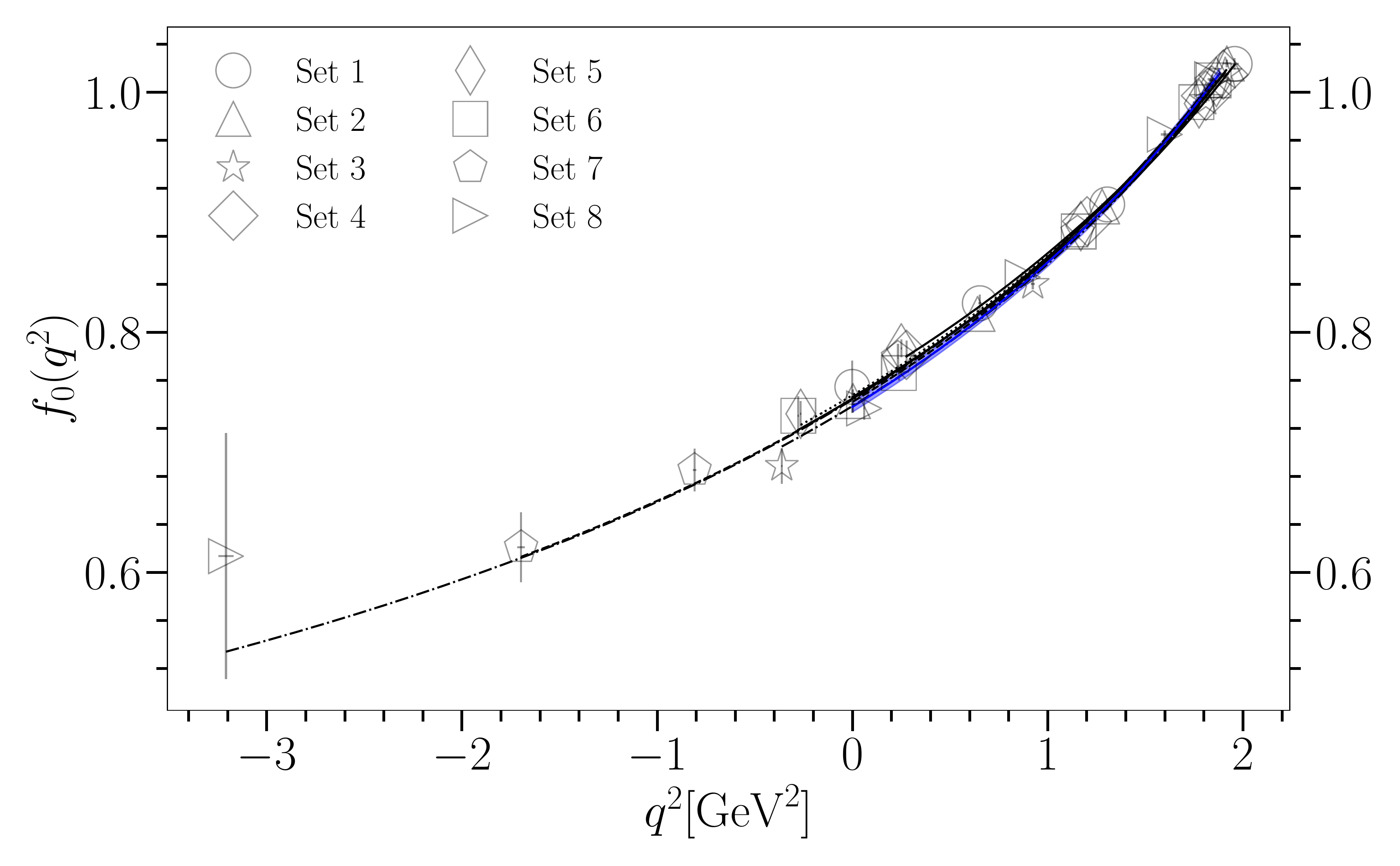}
    \includegraphics[width=0.5\textwidth]{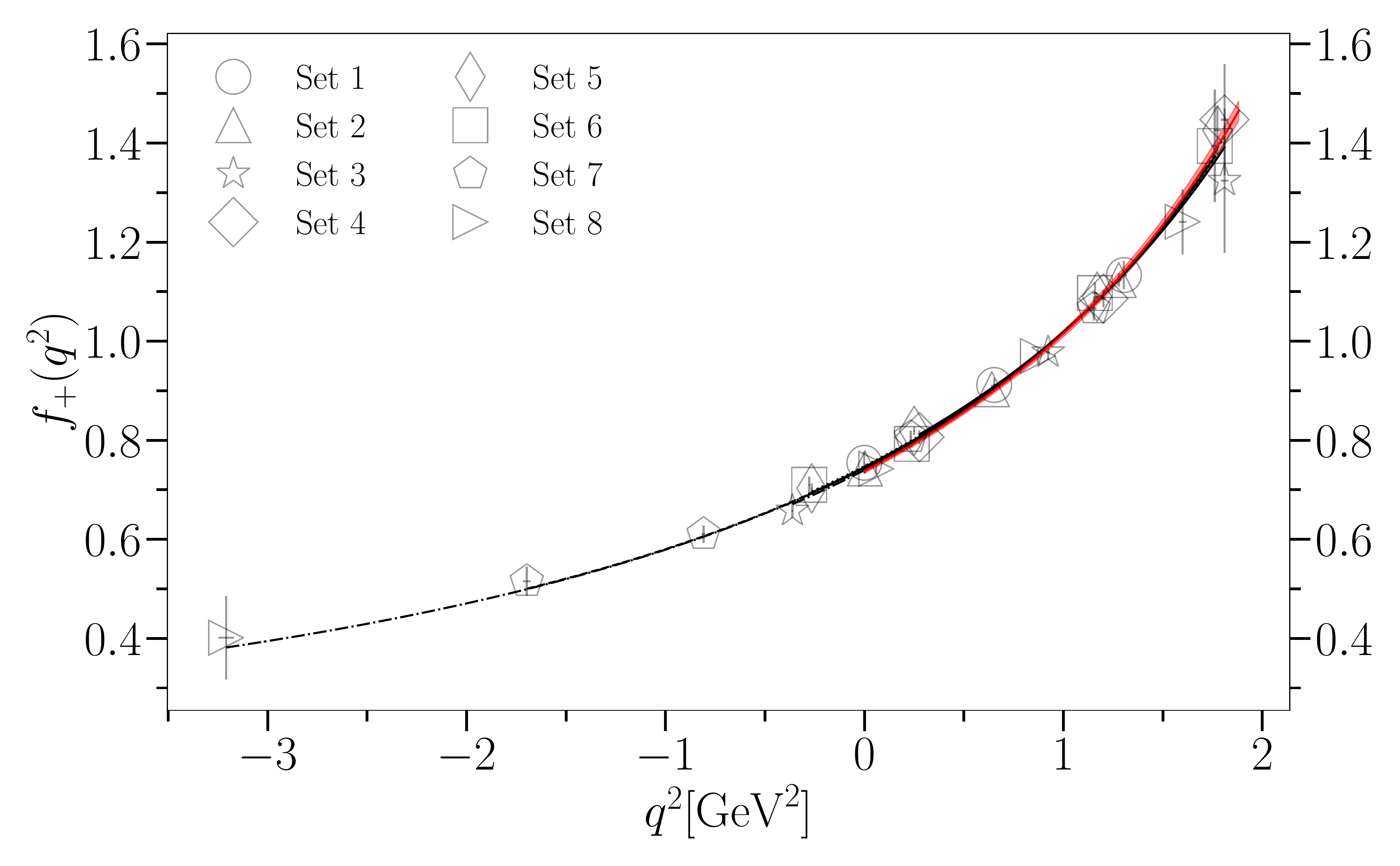}
    \caption{Continuum form factor results for $f_0(q^2)$ and $f_+(q^2)$ (coloured bands), with data from each ensemble shown in black.}
    \label{fig:formfacs}
  \end{center}
\end{figure}

The continuum form factors, as well as the data on each ensemble is shown in Figure~\ref{fig:formfacs}. We see from the similarity of the data across the different lattice spacings that discretisation effects are very small for the HISQ action. 
\subsection{$V_{cs}$}
\begin{figure}
  \begin{center}
    \hspace{-30pt}
    \includegraphics[width=0.5\textwidth]{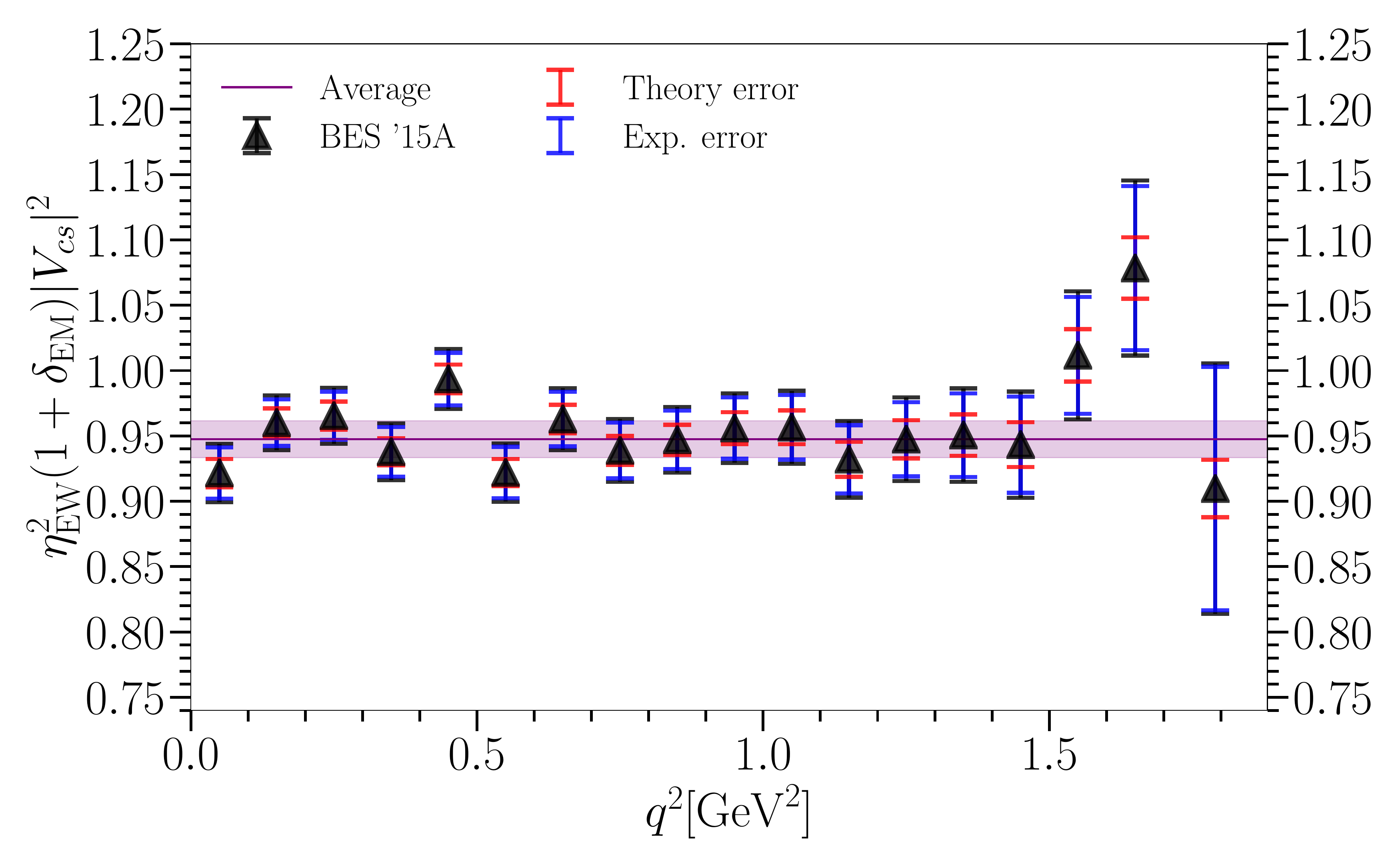}
    \includegraphics[width=0.5\textwidth]{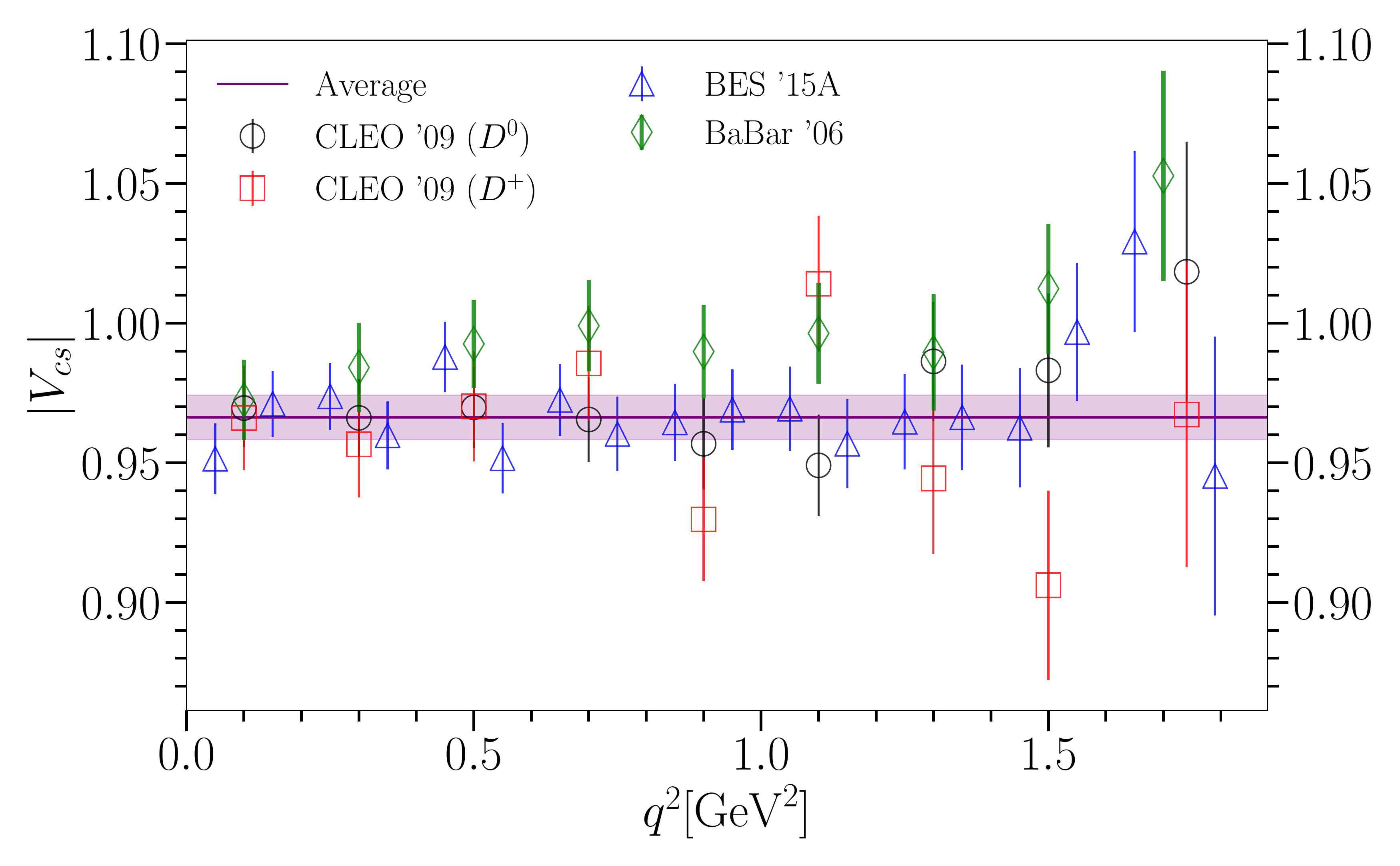}
    \caption{Our preferred method for extracting $V_{cs}$ from binned differential decay rates. Data from one example is show on the left~\cite{Ablikim:2015ixa}, with theoretical and experimental error breakdown. On the right, we show the averages from all four sets of data used in the final analysis~\cite{Besson:2009uv,Aubert:2007wg,Ablikim:2015ixa}.}
    \label{fig:Vcsbins}
  \end{center}
\end{figure}

\begin{figure}
  \begin{center}
    \hspace{-30pt}
    \includegraphics[width=0.5\textwidth]{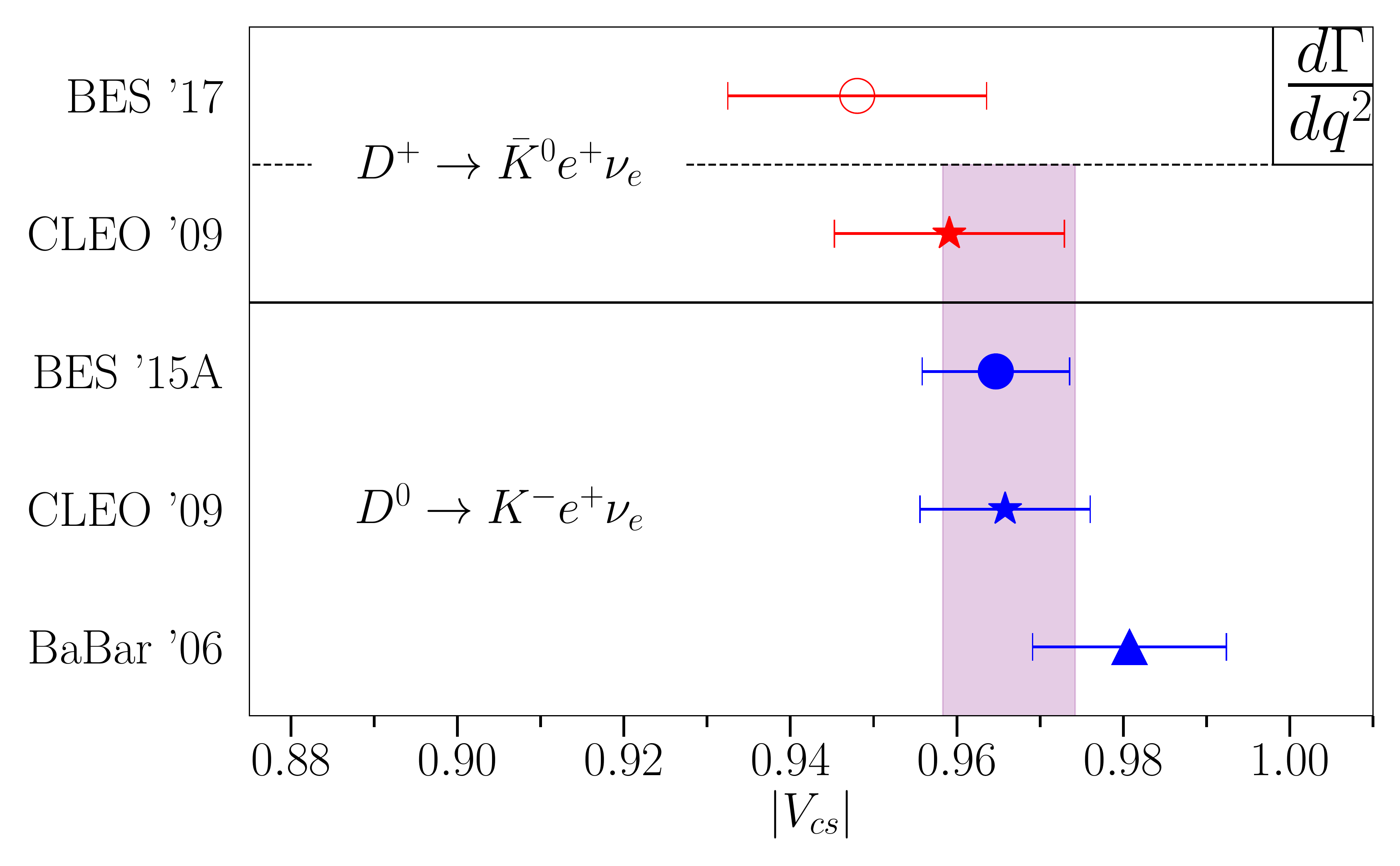}
    \includegraphics[width=0.5\textwidth]{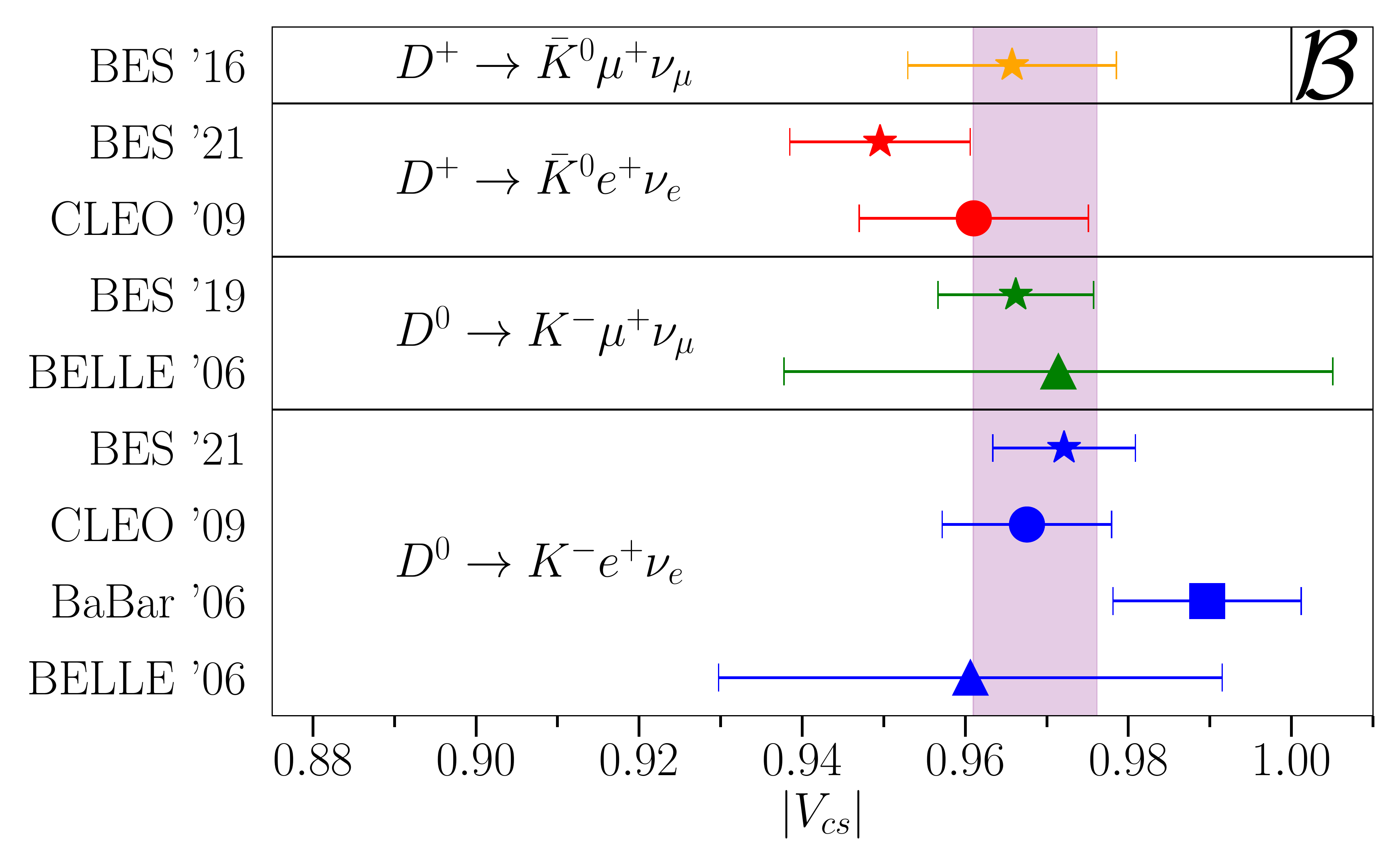}
    \caption{Left hand plot - results from the binned differential decay rate method using experimental data from~\cite{Besson:2009uv,Aubert:2007wg,Ablikim:2015ixa,Ablikim:2017lks}. Right hand plot - results from the branching fraction method using experimental data from~\cite{Ablikim:2016sqt,BESIII:2021mfl,Besson:2009uv,Ablikim:2018evp,Aubert:2007wg,Widhalm:2006wz}. Purple bands show the correlated weighted average.}
    \label{fig:VcsB}
  \end{center}
\end{figure}

With the form factors calculated, we can address the determination of $V_{cs}$. The expression for the differential decay rate is,
\begin{equation}\label{Eq:dgammadq}
  \begin{split}
    \frac{d\Gamma^{D\to K}}{dq^2} =& \frac{G^2_F(\eta_{\mathrm{EW}}|V_{cs}|)^2}{24\pi^3}(1-\epsilon)^2(1+\delta_{\mathrm{EM}})\times\\
    &\Big[|\vec{p}_K|^{3}(1+\frac{\epsilon}{2})|f_+(q^2)|^2 + |\vec{p}_K|M_D^2\Big(1-\frac{M^2_K}{M^2_D}\Big)^2\frac{3\epsilon}{8}|f_0(q^2)|^2\Big],
  \end{split}
\end{equation}
where $\epsilon=m_{\ell}^2/q^2$, for lepton mass $m_{\ell}$ and  $\eta_{\mathrm{EW}}=1.009(2)$, allowing for corrections to $G_F$. Finally, we must allow an uncertainty $\delta_{\mathrm{EM}}$, for final state electromagnetic interactions. This will be larger for the charged kaon case so we allow $0.5\%$ and $1\%$ for the $K^0$ and $K^{\pm}$ respectively. Both $\eta_{\mathrm{EW}}$ and $\delta_{\mathrm{EM}}$ have previously been neglected.\\
Using this expression and our form factors, we can extract $V_{cs}$ in three semi-independent ways. These are semi-independent because they use the same form factors, however, each uses a different combination of experimental data, and a different $q^2$ region. Firstly, our preferred method is to integrate Equation~(\ref{Eq:dgammadq}) over the $q^2$ bins used by experiment, and extract a value for $V_{cs}$ from each bin. This is demonstrated for one set of experimental data~\cite{Ablikim:2015ixa} on the left hand side (LHS) of Figure~\ref{fig:Vcsbins}, whilst the right hand side (RHS) shows the average when all four sets of experimental data are used~\cite{Besson:2009uv,Aubert:2007wg,Ablikim:2015ixa}. We note that experimental uncertainty dominates in each bin, but overall theory error is still larger, as there is more independent experimental data.
Averaging over bins for each experiment gives the values displayed in the LHS of Figure~\ref{fig:VcsB}, where we also include data from~\cite{Ablikim:2017lks}, which was not included in the average as it could not be correlated with other results from the same experiment.

\begin{figure}
  \begin{center}
    \hspace{-30pt}
    \includegraphics[width=0.5\textwidth]{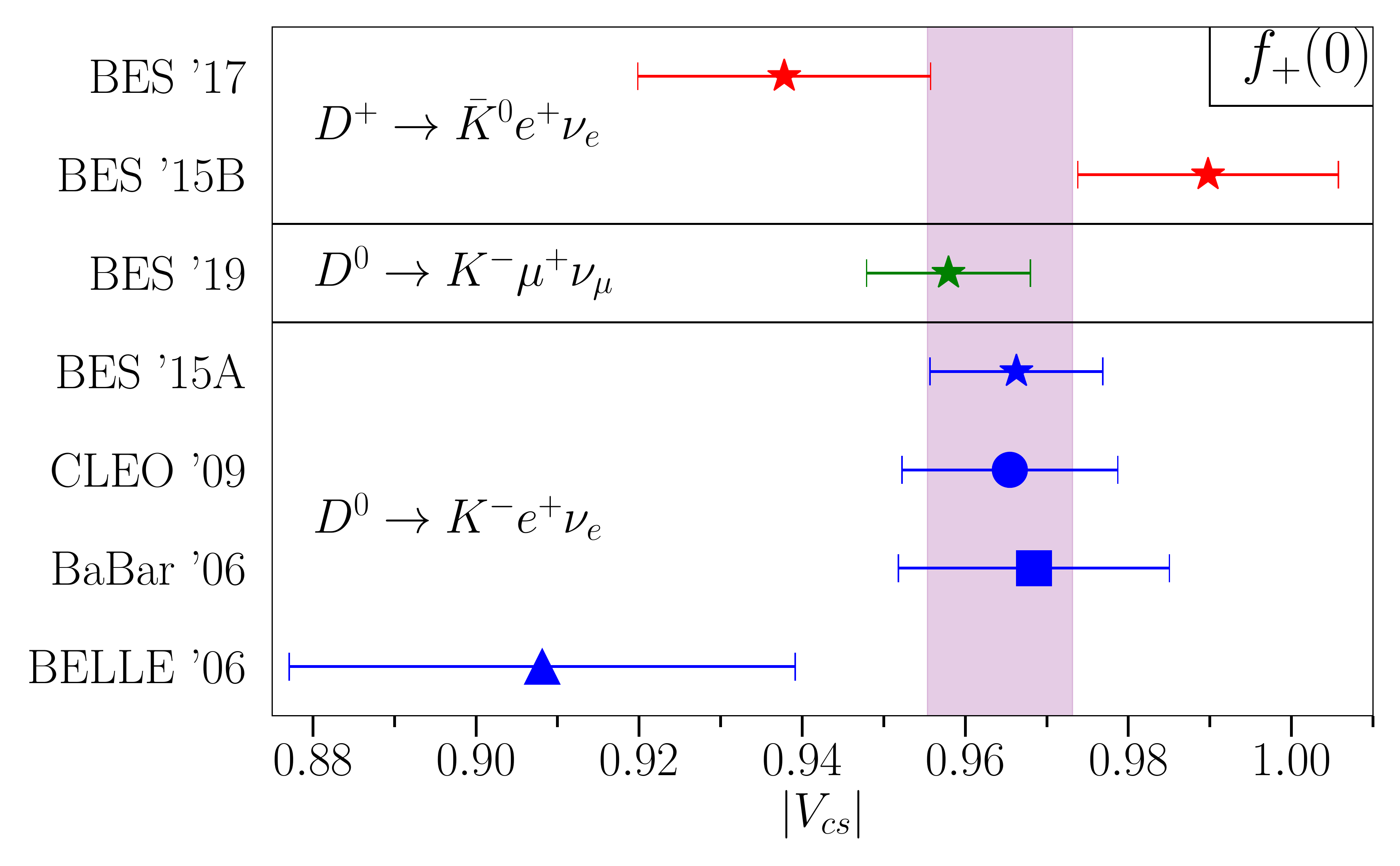}
    \includegraphics[width=0.5\textwidth]{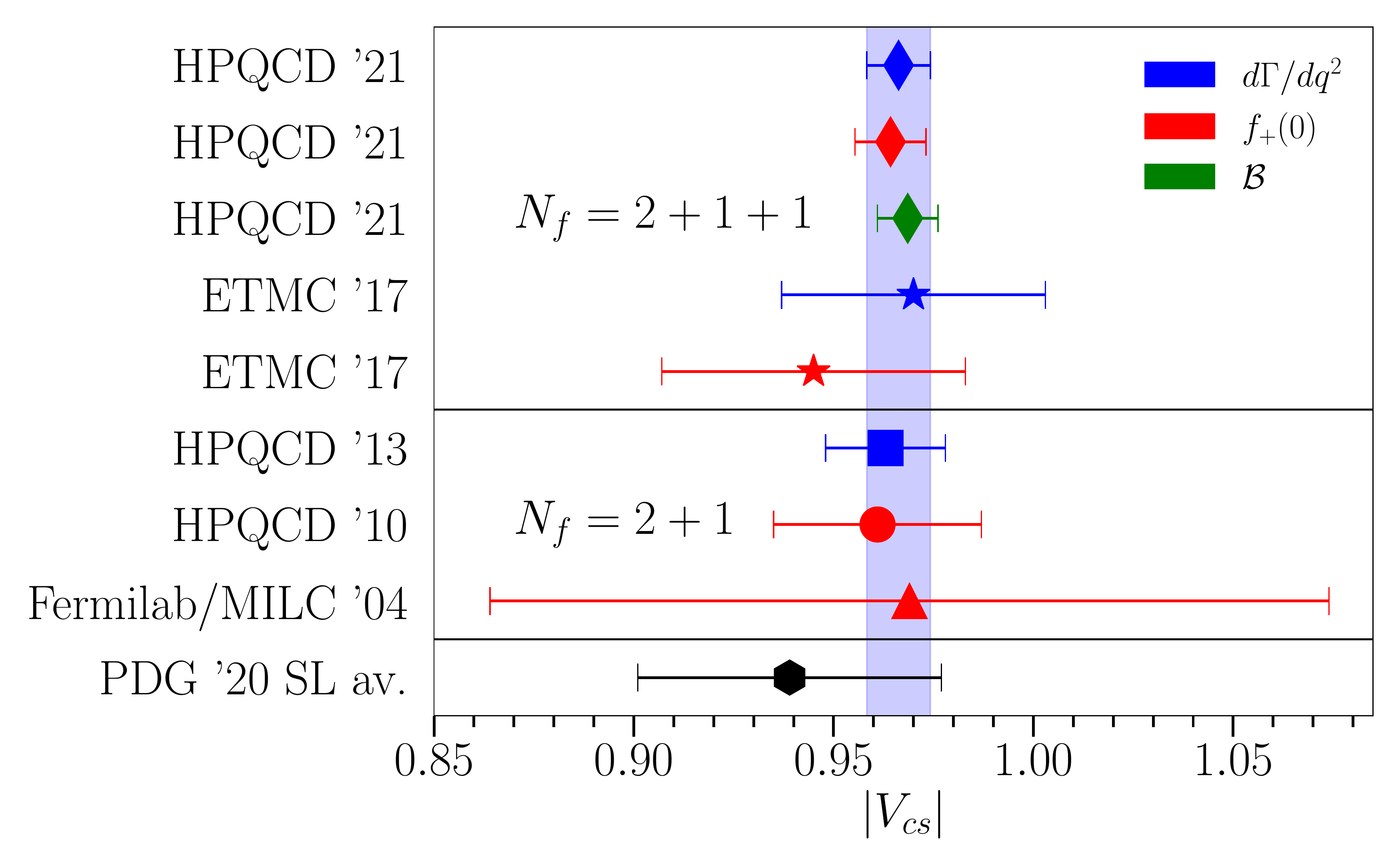}
    \caption{Left hand plot - results from each experiment using the $|V_{cs}|f_+(0)$ method~\cite{Ablikim:2017lks,Ablikim:2015qgt,Ablikim:2018evp,Ablikim:2015ixa,Besson:2009uv,Aubert:2007wg,Widhalm:2006wz}. Right hand plot - comparison of $V_{cs}$ determinations~\cite{Lubicz:2017syv,Riggio:2017zwh,Koponen:2013tua,Na:2010uf,Aubin:2004ej,pdg}; the top three are our new results~\cite{Chakraborty:2021qav}.}
    \label{fig:Vcsf0}
  \end{center}
\end{figure}

The second method we use is to integrate $d\Gamma/dq^2$ across the whole $q^2$ range to obtain the branching fraction $\mathcal{B}$, which can then be combined with a larger set of experiments to extract $V_{cs}$. This is shown on the RHS of Figure~\ref{fig:VcsB}. It's worth noting that this data also accesses the $\mu$ decay channels, which is not the case with our first method.\\
Finally, experimentalists often extrapolate their results to $q^2=0$ and provide a value for $|V_{cs}|f_+(0)$. We can divide this by our $f_+(0)$ value to obtain $V_{cs}$. The result of this is depicted on the LHS of Figure~\ref{fig:Vcsf0}.

The RHS of Figure~\ref{fig:Vcsf0} shows the result of our three methods, as well as previous determinations of $V_{cs}$. We see that our methods agree well, and are much more precise than previous determinations. 
\section{Conclusions}
Our final results for $V_{cs}$ from our three methods are:
\begin{eqnarray}
\label{eq:vcs:list}
&&|V_{cs}|^{\text{d}\Gamma/\text{d}q^2} = 0.9663(53)_{\text{latt}}(39)_{\text{exp}}(19)_{\eta_{EW}}(40)_{\text{EM}}, \nonumber \\
&&|V_{cs}|^{\mathcal{B}} = 0.9686(54)_{\text{latt}}(39)_{\text{exp}}(19)_{\eta_{EW}}(30)_{\text{EM}}, \nonumber \\
&&|V_{cs}|^{{f_+(0)}} = 0.9643(57)_{\text{latt}}(44)_{\text{exp}}(19)_{\eta_{EW}}(48)_{\text{EM}},
\end{eqnarray}
with the first our preferred method. In each case, we see that the theory uncertainty is still the largest contribution, though it is now commensurate with the experimental uncertainty. We also see that there are other non negligible contributions to the uncertainty, most notably from electromagnetic effects, which must be pinned down with further theoretical work if $V_{cs}$ is to be calculated more precisely. Using our final value of $V_{cs} = 0.9663(80)$, the first sub-$1\%$ determination of $V_{cs}$, and the first time it has been shown to be significantly less than $1$, we obtain unitarity constraints,
\begin{equation}
|V_{cd}|^2 + |V_{cs}|^2 + |V_{cb}|^2 =0.9826(22)_{V_{cd}}(155)_{V_{cs}}(1)_{V_{cb}},
\end{equation}
\begin{equation}
|V_{us}|^2 + |V_{cs}|^2 + |V_{ts}|^2 =0.9859(2)_{V_{us}}(155)_{V_{cs}}(1)_{V_{ts}},
\end{equation}
which are consistent with unity. Despite our significant reduction in uncertainty, $V_{cs}$ still dominates the uncertainty.
\begin{figure}
  \begin{center}
    \hspace{-30pt}
    \includegraphics[width=0.5\textwidth]{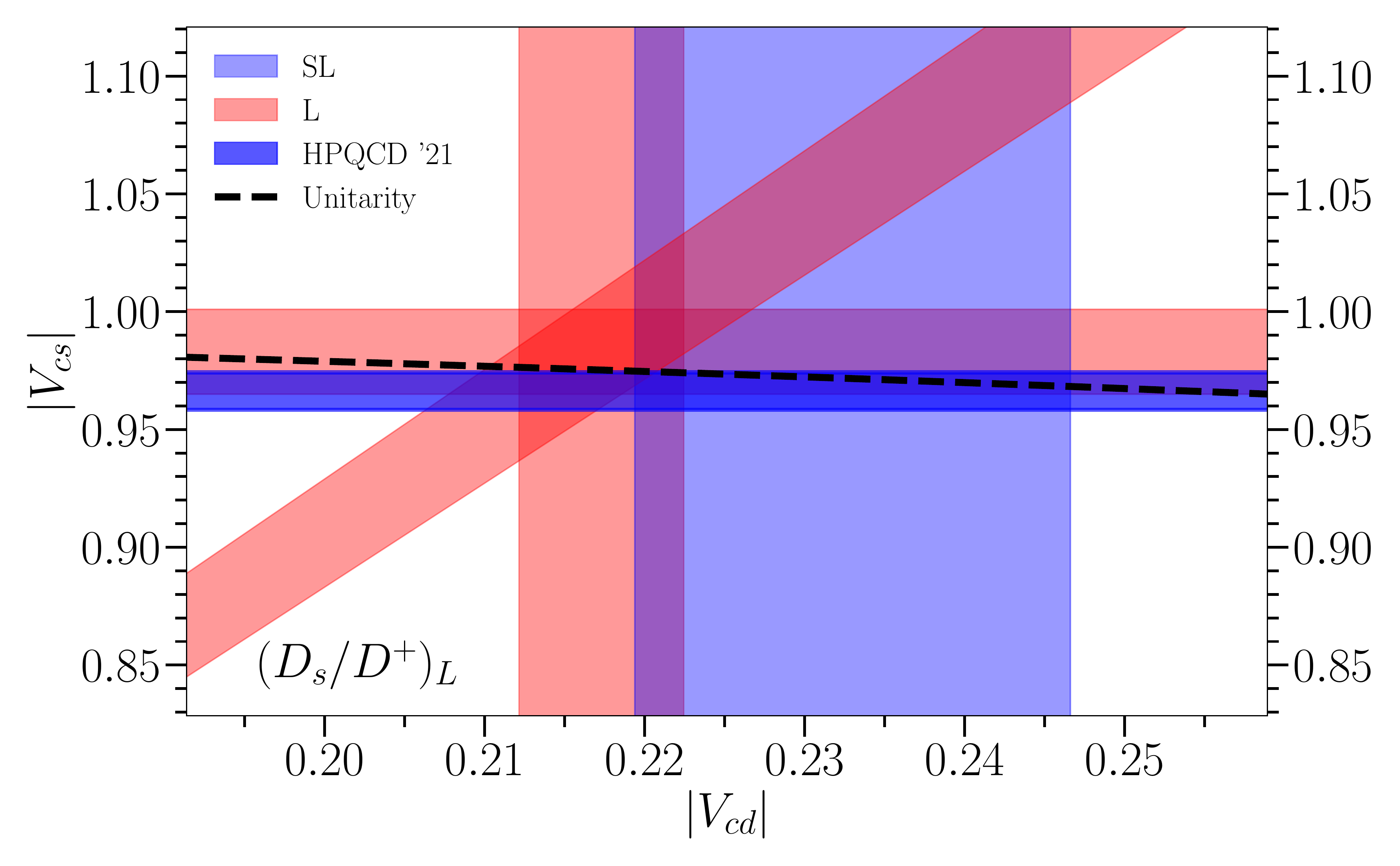}
    \includegraphics[width=0.5\textwidth]{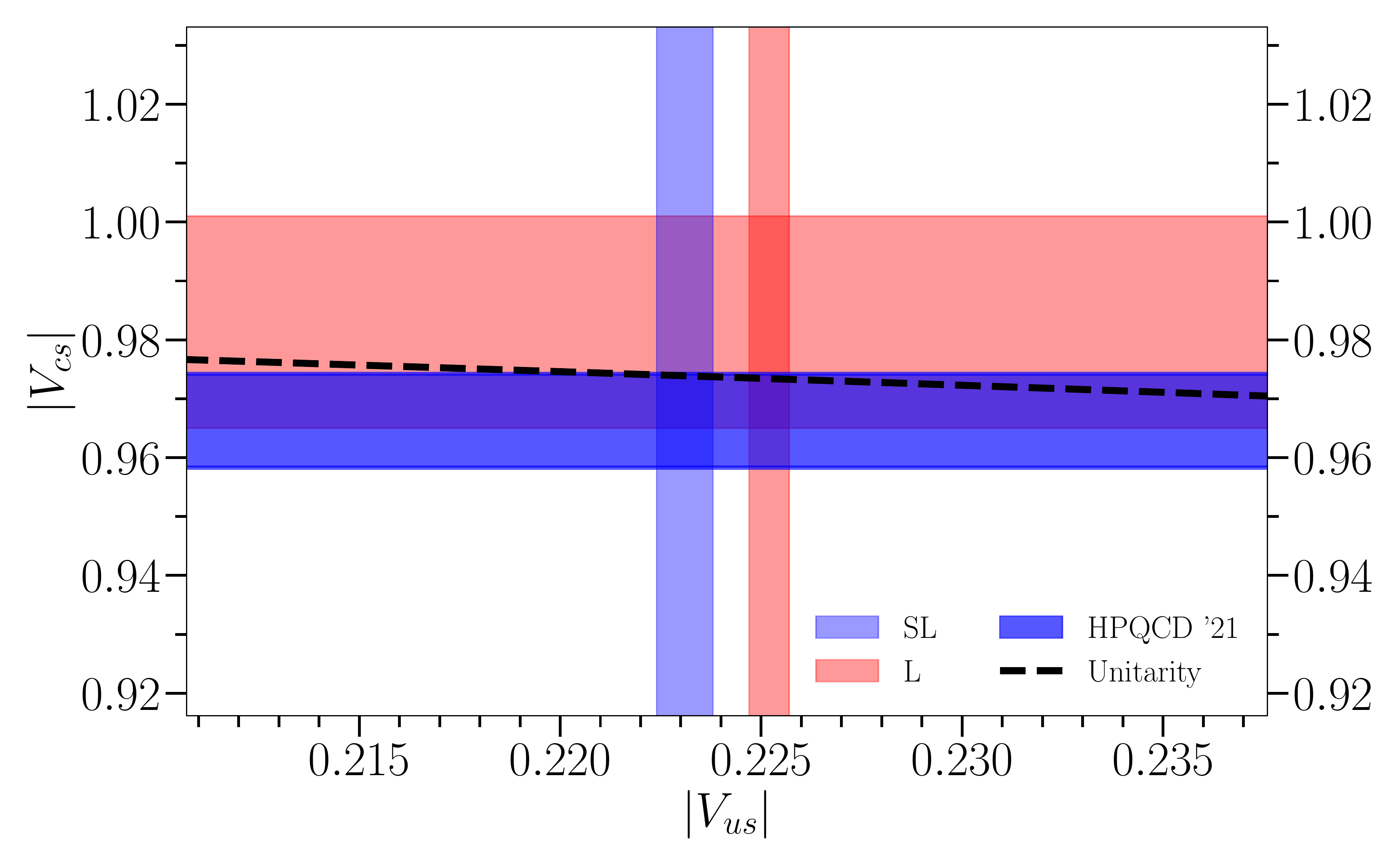}
    \caption{Our semileptonic (SL) $|V_{cs}|$ determination, alongside existing leptonic (L) and semileptonic $|V_{cd}|$ and $|V_{us}|$ values~\cite{pdg}. The black dotted lines represent unitarity, and the diagonal red band on the LHS is a value for the ratio $|V_{cs}|/|V_{cd}|$based on leptonic decays.}
    \label{fig:unit}
  \end{center}
\end{figure}
We show our $V_{cs}$ value in the context of other semileptonic and leptonic determinations and unitarity in Figure~\ref{fig:unit}. We see that our value, which is more precise that the leptonic determination, when combined with unitarity, is unable to distinguish between leptonic and semileptonic $V_{us}$ determinations which are currently in some tension~\cite{Bazavov:2013maa}.
\acknowledgments
We are grateful to the MILC collaboration for the use of their configurations and code. We thank R. Briere, A. Davis, J. Harrison, D. Hatton and M. Wingate for useful discussions. This work used the DiRAC Data Analytic system at the University of Cambridge, operated by the University of Cambridge High Performance Computing Service on behalf of the STFC DiRAC HPC Facility (www.dirac.ac.uk). This equipment was funded by BIS National E-infrastructure capital grant (ST/K001590/1), STFC capital grants ST/H008861/1 and ST/H00887X/1, and STFC DiRAC Operations grant ST/K00333X/1. DiRAC is part of the National E-Infrastructure. We are grateful to the Cambridge HPC support staff for assistance. Funding for this work came from the Gilmour bequest to the University of Glasgow, the Isaac Newton Trust, the Leverhulme Trust ECF scheme, the National Science Foundation and the Science and Technology Facilities Council.
\bibliographystyle{JHEP}
\bibliography{DK}

\end{document}